# AutoDisk: Automated Diffraction Processing and Strain Mapping in 4D-STEM

Sihan Wang[1], Tim Eldred[1,2], Jacob Smith[1], Wenpei Gao[1,2]

[1]Department of Materials Science and Engineering, North Carolina State University, Raleigh, NC 27695
[2]Analytical Instrumentation Facility, North Carolina State University, Raleigh, NC 27695

**Abstract**

Development in lattice strain mapping using four-dimensional scanning transmission electron microscopy (4D-STEM) method now offers improved precision and feasibility. However, automatic and accurate diffraction analysis is still challenging due to noise and the complexity of intensity in diffraction patterns. In this work, we demonstrate an approach, employing the blob detection on cross-correlated diffraction patterns followed by lattice fitting algorithm, to automate the processing of four-dimensional data, including identifying and locating disks, and extracting local lattice parameters without prior knowledge about the material. The approach is both tested using simulated diffraction patterns and applied on experimental data acquired from a Pd@Pt core-shell nanoparticle. Our method shows robustness against various sample thicknesses and high noise, capability to handle complex patterns, and picometer-scale accuracy in strain measurement, making it a promising tool for high-throughput 4D-STEM data processing.

**Introduction**

The structural deformation in materials is closely related to their mechanical, electronic, and chemical properties [1–5]. Strain measurement that probes the structural deformation can offer insights on how the structure is related with materials properties. To determine the strain, methods based on a variety of diffraction and spectroscopic techniques have been developed, including neuron scattering [6], Raman spectroscopy [7], X-ray powder diffraction [8], and X-ray absorption spectroscopy [9]. However, it is still challenging to measure strain and map its distribution in nanosized structures such as those in electronic devices or nanoparticle catalysts, where sub-nanometer spatial resolution is critical [10,11]. To this end, electron diffraction taken using a confined electron probe, performed in a transmission electron microscope (TEM), is promising as it offers both high precision and high spatial resolution in structure characterization, and feasibility to measure strain at the nanometer scale [12–14]. Recently, with the development in four-dimensional scanning transmission electron microscopy (4D-STEM), a technique that uses high speed electron cameras to capture two-dimensional diffractions as the electron probe is raster scanned on the sample [15–17], the advantage of electron diffraction method has become more obvious. By measuring reciprocal lattice using the diffraction pattern, the structure and strain in each probe position can be determined and mapped [18,19]. Compared with direct strain measurements on images taken in real space [20], including geometric phase analysis [13] and peak pairs analysis [20,21], 4D-STEM can be applied over a large area, on a variety of materials, and with less constraint in the local crystal orientation [16]. As a result, strain mapping using 4D-STEM has now been widely applied in electronic devices [22], structural materials [23], in-situ deformed samples [24,25], and two-dimensional materials [26].

In the 4D-STEM method, the accuracy in locating diffraction disks determines the ultimate precision in strain measurement. Positions of the diffraction disks are also the basis of other analyses such as symmetry and phase mapping, which rely on quantifying the intensity and its distribution in individual diffraction disk. Many algorithms have been developed to identify disks in a diffraction pattern, including cross-correlation based algorithms using a different filters [27–31], edge detection [30], exit-wave power cepstrum based method [32], and template matching approaches with patterned probe imprinted in the TEM aperture [33]. After the diffraction disks are detected, peak fitting by center of mass (CoM) and two-dimensional (2D) Gaussian-fitting algorithms can be easily applied to identify disk centers [26,31,32]. Circular Hough transform can also be applied to extract circular features during disk position detection [11,34], which performs well on locating disks, though sensitive to background noise. In spite of these technical development, diffraction analysis in 4D-STEM remains challenging due to the following reasons. First, with the improved acquisition rate in 4D-STEM, data sets of more than 100 thousand diffraction patterns can be recorded within minutes, with typical data sizes of tens to hundreds of GB, rendering any time-consuming manual data-labeling not feasible. Within each diffraction pattern, the intensity variation in individual diffraction disks greatly limits the accuracy to locate the disk centers, especially when the diffraction disks are large. To improve the accuracy, optimization method such as circle fitting has been developed but hindered by the expensive computational cost [28]. These challenges are exacerbated when 4D-STEM is applied to beam-sensitive materials such as 2D structures, metal–organic frameworks, and zeolites. An automatic, accurate, and robust approach for diffraction analysis is necessary.

Here, we show a workflow to automatically determine diffraction disk positions from diffraction patterns in 4D-STEM, compute local lattice parameters, and map strain distributions with high precision. The automatic disk registration (AutoDisk) method detects diffraction disks in each independent pattern, refines their positions to minimize errors due to intensity variation, and optimizes the position estimation by fitting these disks with a 2D lattice. Local lattice parameters and distortion in real space can then be measured based on the information on the reciprocal lattice. Meanwhile, this entire approach requires neither prior knowledge of the material structure nor any manual intervene to measure the local lattice parameters from a 4D data set, therefore one noteworthy advantage is that the automation can avoid inconsistency due to human interactions in strain mapping methods. The precision of AutoDisk and its robustness against noise are demonstrated by applying AutoDisk to both simulated diffraction patterns using Bloch wave method [35] and by mapping the strain of a Pd@Pt core-shell octahedral nanoparticles.

**Method**
*Experiment*
In 4D-STEM (**Fig. 1a**), an electron probe is rasterized across the specimen, convergent beam electron diffraction (CBED) patterns form and are captured at the back-focal plane. The 4D-STEM experiment is performed in an aberration corrected Thermo Scientific Titan STEM operated at 200 kV. CBED patterns were acquired using an EMPAD [36] detector. The 4D data sets consist CBED patterns from all probe locations. In strain mapping of the Pd@Pt octahedral nanoparticles (**Fig. 1b**), a calibrated semi-convergence angle of 4.26 mrad is used, corresponding to a probe size of 0.9 nm in diameter. A typical 4D data set, taken with the nanoparticle tilted along the Pd[110] zone axis, contains $256 \times 256$ diffractions, corresponding to an area of $59 \times 59$ nm in size. The CBED patterns are recorded with $128 \times 128$ pixels in frame size at the acquisition rate of 1000 frames/second (fps). In the 4D data set, each CBED pattern contains a center disk (000) and diffraction disks with the indices shown in **Fig. 1c**. The reciprocal lattice in the diffraction pattern can be represented by two basis vectors from the center disk to a pair of non-colinear first-order disks; here, the two basis vectors $\vec{a}$ and $\vec{b}$ are labeled as (000) to ($1\bar{1}1$) and ($1\bar{1}\bar{1}$).

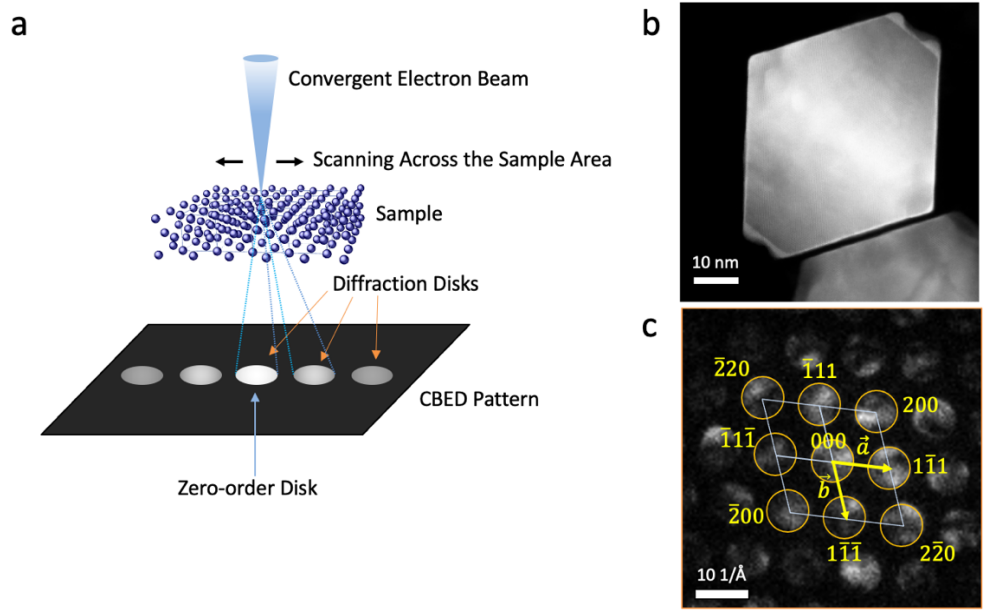

**Fig. 1.** (a) A schematic of 4D-STEM. (b) A high-angle annular dark-field (HAADF) STEM image of a Pd@Pt core-shell nanoparticle. (c) The indexed diffraction pattern from the nanoparticle in (b) with two basis lattice vectors marked as $\vec{a}$ and $\vec{b}$ used for diffraction analysis.

*Diffraction Simulation*

      CBED patterns of Pd FCC along Pd[110] were simulated using Bloch wave method [37]. In general, intensity variation inside diffraction disks is more pronounced with higher thickness and larger convergence angles. To test how sample thickness and noise influence the data analysis, CBED patterns with varying sample thicknesses from 1 nm to 40 nm were simulated, with different levels of Poisson and Gaussian noise added.

*Diffraction Analysis*

      Our approach for data analysis is implemented in python and all the source code, as well as the sample data, are released on GitHub. Functions from packages including NumPy [38], OpenCV [39] and Matplotlib [40] were used to read, process and visualize the images and results. Scikit-image [41], an open-source library with a collection of algorithms for image processing, was applied for methods used in disk detection. A demo tutorial is also provided as a Jupyter Notebook to illustrate each step in the process.

**Automation in Diffraction Analysis**

      In general, the automated diffraction analysis includes disk detection, refinement, reciprocal lattice fitting and optimization. Diffraction disks in a CBED pattern have a size and shape defined by the probe forming aperture in STEM. Lens aberration and structure distortion of the sample may cause shape distortion in high-order diffraction disks that are away from the optical axis. Thus, to eliminate the error caused by the distortion in the diffraction patterns on pattern

identification, the shape and radius of the diffraction disk are determined from the center disk of a pattern taken over vacuum or an average of a stack of patterns. The edge of the center disk is detected by taking the maximums of the second derivative of its intensity, the detected radius of this disk edge is recorded as $r_0$. All the diffraction disks in the same dataset are identified and located initially based on this shape and radius. After the disks are first identified, we fit the positions of the disks with a 2D lattice defined by two basis vectors, which are those from the zero-order disk to its nearest first-order disks. Lattice parameters in the reciprocal space are then estimated by optimizing basis vectors. Finally, the local lattice parameter in real space is calculated based on the reciprocal lattice measured at each probe position. The deviation of local lattice from a reference lattice is utilized to map the normal and shear strain. Each step in this analysis is detailed in the following sections.

*Automated Diffraction Disk Detection*

The diffraction analysis is started with disk detection, in which the number and positions of diffraction disks on a CBED pattern are estimated. To improve visibility of the disk edges, a new pattern is generated using the square root of the intensities. Gaussian blur can be applied to denoise the images if the background is noisy. Then we cross-correlate the sub-areas of the diffraction pattern with a ring filter, as shown in **Fig. 2a,** to transform each diffraction disk into a plate with a bright center blob (**Fig. 2b**) in the cross-correlation map. The new map will be used later to determine the center of each diffraction disk. Here, to reduce the detection error caused by skewed intensities along the disk edges, the radius of the ring filter $r_f$, is chosen to be slightly smaller than $r_0$, the radius of the center disk, as represented in **Eqn. 1**,

$$r_f = c * r_0 \qquad (1)$$

where $c$ is a coefficient between 0 and 1. The selection of $c$ affects the size of the blob feature in the cross-correlated pattern, a value slightly lower than 1 used in the ring filter would make the center area of each disk relatively uniform, especially when large intensity variation exists inside the diffraction disks. By comparing the disk detection results (**Fig. 3c**) with those derived directly from the diffraction pattern without the ring filter (**Fig. 3b**), we show applying the ring filter can greatly improve the accuracy of the disk detection.

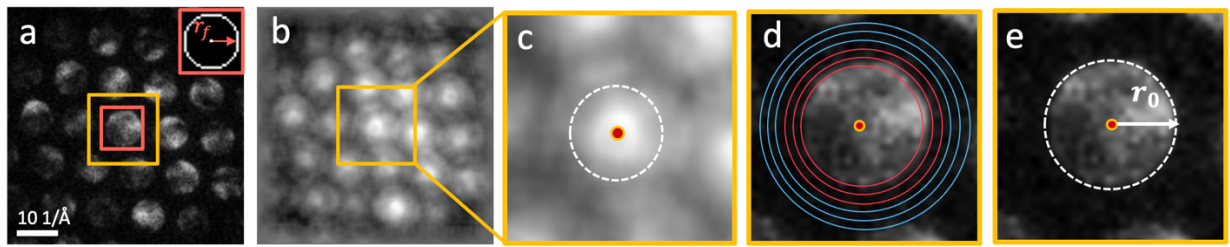

**Fig. 2.** The schema of the disk center detection. (a) The original CBED Pattern and the ring filter with radius $r_f$ used for cross-correlation. (b) The cross-correlated pattern of (a). (c) A detected center blob with LoG filters in the selected area of (b). (d) RGM method applied on the detected disk with red and blue circles as interior and exterior circles, respectively. (e) The refined center of one disk and the rim of the disk with radius $r_0$, marked in the white dashed curve.

After cross-correlation, we apply a blob detection with Laplacian of the Gaussian (LoG) filters to calculate the coordinates of the disk centers [42,43]. The LoG filter is a feature extractor in computer vision, which is widely used for blob detection, written as **Eqn. 2**. The *x* and *y* are pixel coordinates and $\sigma$ is the scale related to $r_0$ defining the size of the blobs. Specifically, each cross-correlated pattern is convolved with scale-normalized LoG filters to build a scale-space representation. The Laplacian is maximum for the circle of radius at about $0.5r_0$, a half of the estimated radius of the center blobs. The local maxima on the filtered pattern represents the detected blob centers, and the coordinate of each center is used as our first estimate of the center position of each diffraction disk, shown in **Fig. 2b** and **Fig. 2c**. Then, a list of coordinates of disk center positions is detected and stored in the set of *$p_1$* from this step.

$$LoG(x,y) = -\frac{1}{\pi\sigma^4}\left[1 - \frac{x^2+y^2}{2\sigma^2}\right]e^{-\frac{x^2+y^2}{2\sigma^2}} \qquad (2)$$

For a diffraction pattern consisting of diffraction disks with strong uniform internal intensity, the disk locations in *$p_1$* already have an accuracy at the picometer-scale. However, the accuracy can be influenced by noise and intensity variation inside the diffraction disks. At high noise level, the center detection can have huge errors. Also, with large intensity fluctuation in a diffraction disk, the detected center would shift from the geometric center to the area with higher intensities. **Fig. 3b** and **3c** show a comparison between a LoG blob detection on the original pattern and on its cross-correlated pattern. As mentioned above, the cross-correlation using a ring filter can improve the detection accuracy. However, the disk positions detected in **Fig. 3c** still show some noticeable errors. To improve the accuracy, a further refinement is necessary.

To account for the effect of the intensity variation and the background noise, we adapt the basic algorithm behind the radial gradient maximization (RGM) method developed by Müller et al [28] and modified it to better work with diffraction disks with intensity variation. This method distinguishes disks from the background by maximizing the intensity gradient in the radial direction of each detected disk and has been shown to improv on the accuracy in detecting the disk centers [28,30,44]. Therefore, after the blob detection, candidate pixel coordinates within a square window of $M \times M$ pixels in sizes around each position in *$p_1$* are assigned as the centers of concentric rings with radius from $0.8r_0$ to $1.2r_0$, illustrated in **Fig. 2d**. To calculate the radial gradient, we subtracted the sum of the mean intensities of the exterior circle with radius from $1.0r_0$ to $1.2r_0$ from that of each interior circle with radius from $0.8r_0$ to $1.0r_0$, and save it as weight *w*. In most cases, the higher the radial gradient is, the closer the chosen coordinate is to the true center of the diffraction disk. A new center coordinate set *$p_2$* therefore replaces *$p_1$* after this refinement. With high noise, a threshold related to the signal-to-noise ratio (SNR) of the original pattern can be used to filter out wrongly detected disks. Only when the radial gradient of a detected disk exceeds the threshold, this corresponding diffraction disk would be recognized.

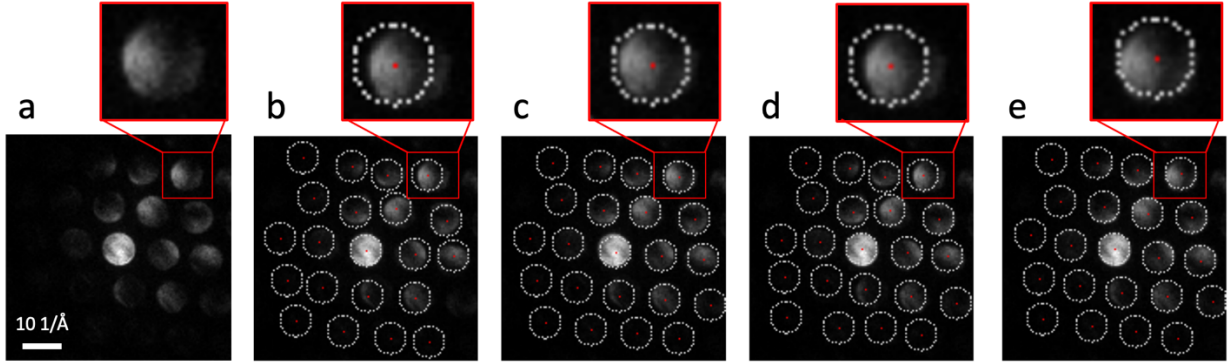

**Fig. 3.** A comparison of the accuracy of disks detection using different methods. (a) An experimental CBED pattern from the Pd@Pt nanoparticle. (b) Disk positions determined by LoG blob detection on the original diffraction pattern. (c) Disk positions detected by the center blob detection on the cross-correlated pattern. Based on the disk positions estimated in (c), (d) and (e) show the refined positions by CoM and the modified RGM method, respectively.

After refinement using RGM, the accuracy is further improved (**Fig. 3e**), compared with results using only blob detection in **Fig. 3c**. We also compared CoM [23,26,31], another commonly used refinement in peak detection, with RGM. Results in **Fig. 3d** are CoM refined disk positions. It's clear that with large diffraction disks, RGM refinement outperforms CoM.

*Lattice Fitting*

The positions of disks in a diffraction pattern from a single crystal are on a two-dimensional (2D) lattice, which is the 2D projection of the 3D reciprocal lattice of the crystal. The 2D lattice can be represented by two non-parallel basis vectors. Here a lattice fitting algorithm is designed to automatically fit the positions of the detected disks in a 2D reciprocal lattice. With calculation and refinement, the corresponding lattice in the real space can then be determined with high precision.

To define a 2D lattice, two basis vectors ($\vec{a}$ and $\vec{b}$) are determined in a diffraction pattern. As shown in **Fig. 4**, the vector from the center disk to its nearest neighbor is firstly defined as the basis vector, $\vec{a}$ (**Fig. 4b**) and the coordinates of the detected disks in $p_2$ are then transformed from the default coordinate system ($X_0$, $Y_0$) to a new system ($X_1$, $Y_1$) centered at the center disk to align $\vec{a}$ with the $X_1$ direction, which is also defined as the *horizontal* direction in the later analysis. The disk positions represented in the new coordinate system are now $p_3$.

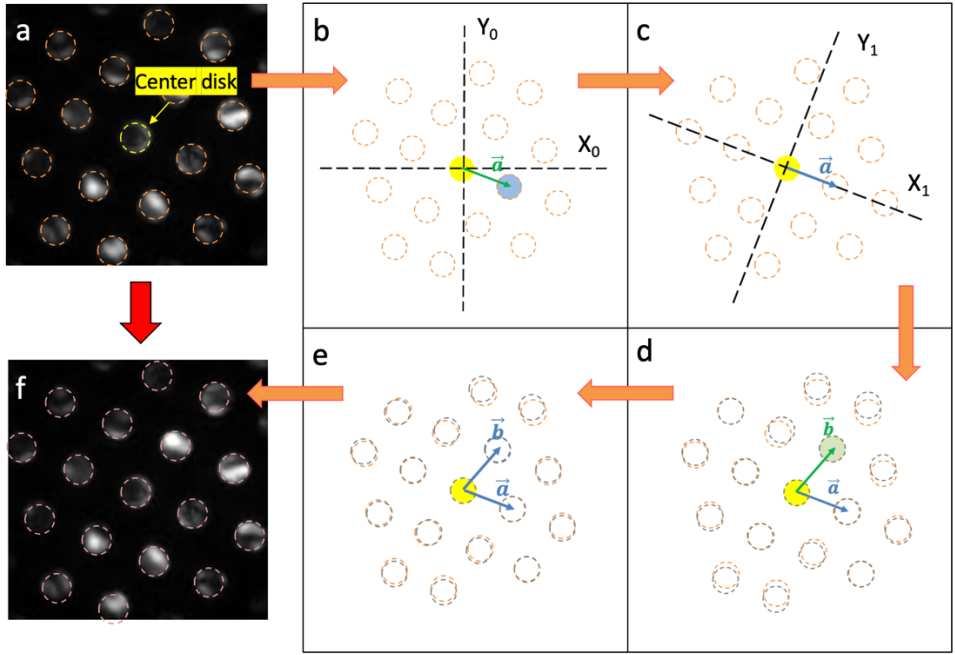

**Fig. 4.** A scheme of the lattice fitting algorithm. (a) All detected disk positions in $p_2$ are labeled in orange on the original diffraction pattern except the center disk marked in yellow. (b) $\vec{a}$ is defined as the vector from the center disk (colored in yellow) to its nearest neighboring disk (colored in blue). (c) All coordinates are transformed so that $\vec{a}$ is along the $X_1$ direction in the new coordinate system, and $\vec{a}$ is refined as the average distances between two horizontally neighboring disks. (d) The vector from the center disk to its non-horizontal nearest neighbor (colored in green) is assigned as the initial $\vec{b}$ and a hypothetical lattice is generated and labeled in black. (e) $\vec{b}$ is optimized by minimizing the deviations between the artificial lattice points and the detected disk positions. (f) The final disk positions labeled on the original diffraction pattern.

After the coordinate transformation, $\vec{a}$ should have a non-zero $x$ component while its $y$ component being 0 (**Fig. 4c**), represented as $\vec{a}$ ($x_a$, 0). Here, $x_a$ should be the average Euclidean distance between all horizontally neighbored diffraction disks. Therefore, all coordinates in $p_3$ are grouped based on their proximity in $y$ components. Positions with similar $y$ are in the same group. After grouping, a refinement on the rotation angle $\bar{\theta}$ is performed using **Eqn. 3** where $n$ is the total number of horizontally neighboring pairs and ($\Delta x$, $\Delta y$) is the corresponding vector from the left point to the right one. We then rotate the coordinates of disks in $p_3$ by $\bar{\theta}$ and update the coordinates to $p_4$.

$$\bar{\theta} = \frac{1}{n}\sum_{i=1}^{n} tan^{-1}(\Delta y / \Delta x) \tag{3}$$

From $p_4$, a weighted average of the y values of each group is calculated using **Eqn. 4**, with the weights, $w$, being the radial gradient calculated from the previous RGM refinement. For example, in the group $k$, the $y$ values of its members from 1 to $n$ are assigned to $y_k$, $w_{ki}$ is the weight of each position in the group.

$$\tag{4}$$

$$y_k = \frac{1}{\sum_{i=1}^{n} w_{ki}} \sum_{j=1}^{n} w_{kj} \cdot y_{kj}$$

At the same time, the lattice vector $\vec{a}(x_a, 0)$ is refined using the average spacing between every neighboring position in the horizontal direction. If there is a missing lattice point, the distance of two neighboring disks close to that point would be far greater than the average; if there is a wrongly detected disk, for example, due to high noise, the distance between its neighbors would also be far off. Thus, only the spacing with values between the two limits defined by the Tukey's rule [45] are counted, represented in **Eqn. 5** where $C_U$ and $C_L$ are the upper and lower limits respectively, while $Q1$ and $Q3$ are the first and the third quantiles (25th and 75th percentiles) of the distance values. Then, since the horizontal lattice vector has only the $x$ component, the average spacing between each pair of neighbors is the magnitude of the first basis vector $\vec{a}$.

$$\begin{aligned} C_U &= Q3 + 1.5 * (Q3 - Q1) \\ C_L &= Q1 - 1.5 * (Q3 - Q1) \end{aligned} \quad (5)$$

The second basis vector $\vec{b}(x_b, y_b)$ of the lattice is estimated using a vector from the center disk to its next nearest neighboring disk having a non-zero $y$ component, as shown in **Fig. 4d**. To optimize $x_b$ and $y_b$, a reference lattice of disk positions, is generated using $\vec{a}$ and $\vec{b}$, and the coordinates of the reference lattice points, $(x_i', y_i')$, are compared with that of the detected disk positions, $(x_i, y_i)$ where $i$ is the index of each detected disk center. The values of $x_b$ and $y_b$ are refined by minimizing the deviances between the lattice points $(x_i', y_i')$ and the detected centers $(x_i, y_i)$, written as **Eqn. 6** for all detected disks, where $n$ is the total number of detected disks and *argmin* is the argument to minimize the function values, illustrated in **Fig. 4e**.

$$\begin{aligned} y_b &= argmin\left\{\frac{\sum_{i=1}^{n}(y_i - y_i')^2}{n}\right\}, \\ x_b &= argmin\left\{\frac{\sum_{i=1}^{n}(x_i - x_i')^2}{n}\right\}. \end{aligned} \quad (6)$$

The optimized $\vec{a}$ $(x_a, 0)$ and $\vec{b}$ $(x_b, y_b)$ are eventually transformed to the pre-defined coordinate system to represent the 2D lattice formed by the diffraction disks. The 2D lattice in the diffraction pattern corresponds to the atomic structure at the electron probe position. Applying this method to the entire 4D data set results in the map of local lattice for all probe locations, allowing for further analysis of strain and structure distortion.

**Results and Discussion**
*Robustness against Noise and Change in Sample Thickness*
    Both sample thickness and noise may affect the accuracy in analysis of the diffraction data, here we test the robustness of AutoDisk using simulated diffraction patterns. The CBED patterns of Pd along its [110] zone axis are simulated using Bloch wave method with the sample thicknesses from 1 nm to 40 nm. Gaussian noise and Poisson noise are added to the simulated patterns. AutoDisk is applied to analyze the simulated data. The basis lattice vectors $\vec{a}$ and $\vec{b}$ determined automatically by AutoDisk are $(1\bar{1}1)$ and $(1\bar{1}\bar{1})$, respectively.

The error is computed as the deviation of the measured lattice parameters from the input structure used in the simulation. The SNR is defined as the ratio of the mean intensity in the center disk to the standard deviation of the background in the diffraction pattern, represented in **Eqn. 7**, which is also used by Yuan et al. [11].

$$SNR = \frac{\overline{I_{Center\ disk}}}{\sigma_{Background}} \tag{7}$$

In **Fig. 5a-h**, the disks identified by AutoDisk are labeled on the set of CBED patterns with sample thickness of 40 nm and SNR values from infinity (no background noise) to 2.4. In each pattern, the basis vectors are converted to the magnitude of the estimated d-spacing in real space, and the error is calculated.

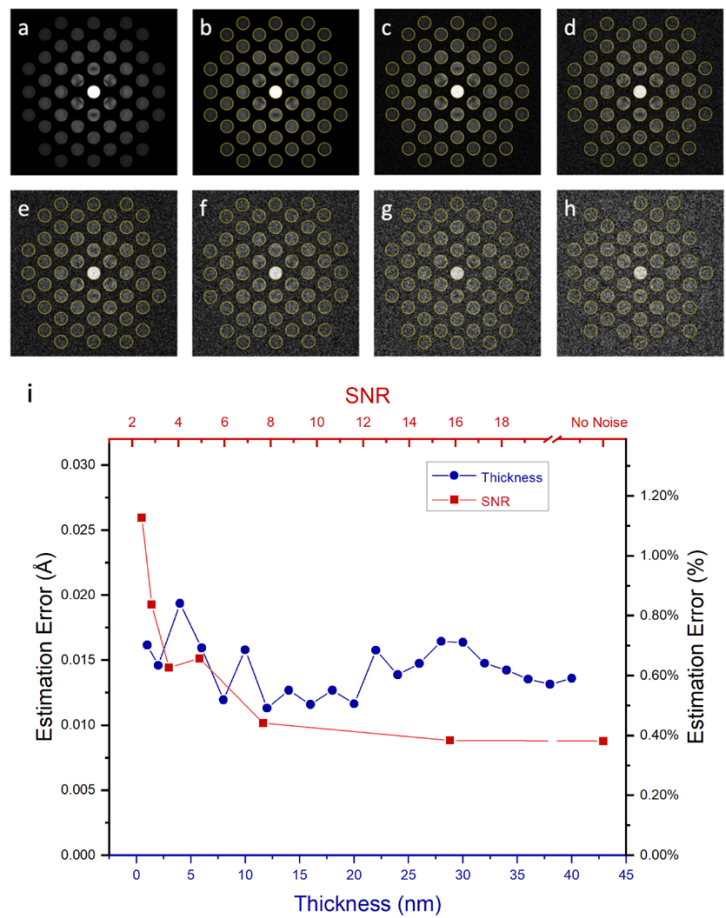

**Fig. 5.** Disk registration and error analysis on simulated CBED patterns. (a-h) Results of disk registration from simulated diffraction patterns with different noise level: (a) The simulated CBED pattern (b) SNR = inf, i.e. no background noise and no Gaussian noise present, (c) SNR = 15.8, (d) SNR = 7.7, (e) SNR = 4.9, (f) SNR = 3.6, (g) SNR = 2.8, (h) SNR = 2.4, (i) Errors of the estimated lattice parameter in real space with respect to different thicknesses and SNRs.

Without noise, no obvious dependency of the error is seen on the sample thickness, and AutoDisk keeps a detection error of $0.0086 \pm 0.002$ Å. With the increase of noise, the error rises from 0.01 Å to 0.026 Å. Note that even with high noise, for example, at SNR = 2.4 (**Fig. 5h)**, most disks can be correctly detected in the diffraction, offering sufficient measurement accuracy.

*Strain Mapping*

AutoDisk is applied to experimental 4D STEM data set for a Pd@Pt nanoparticle oriented along the [110] zone axis. The *X* and *Y* axes defining the coordinates are along the (2$\bar{2}$0) and (002) labeled in the diffraction pattern in **Fig. 6a**, the corresponding directions in real space are also marked in the bright-field (BF) image in **Fig. 6b** and the reconstructed annular dark-field (ADF) image in **Fig. 6c**. Disk locations identified by AutoDisk on some example patterns are shown in **Fig. 7b-i**, and their corresponding positions in the nanoparticle are marked in **Fig. 7a**. Results show that AutoDisk works in most scenarios, regardless of the brightness and uniformity of the intensity in the diffraction patterns.

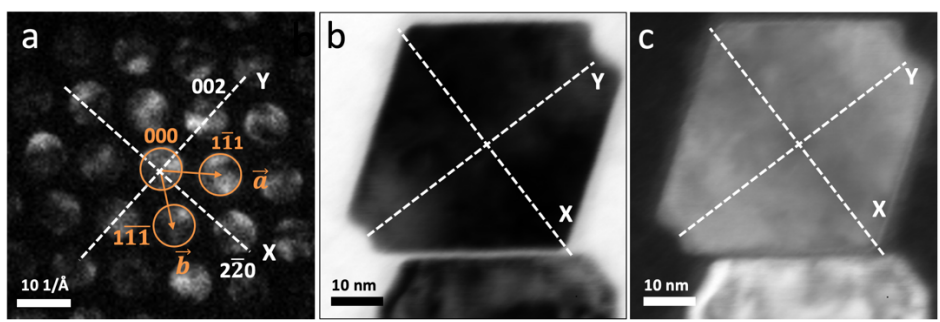

**Fig. 6.** Local coordinate system used for strain measurement. (a) Lattice vectors $\vec{a}$ is from (000) disk to (1$\bar{1}$1) disk and $\vec{b}$ from (000) disk to (1$\bar{1}\bar{1}$) disk. The *X*-axis is along (000) to (220) and *Y*-axis along the (002). (b-c) The corresponding *X* and *Y* axes are marked in the BF and reconstructed ADF image.

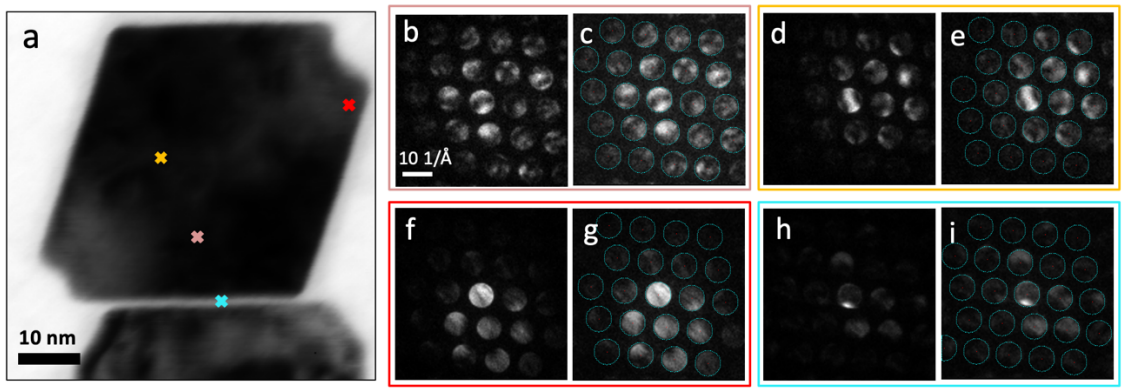

**Fig. 7.** CBED patterns and the identified diffraction disks. (a) The BF STEM image of the nanoparticle, with the positions of representative CBED patterns marked in small crosses, (b-c) a typical pattern with most diffraction disks visible and having clear edges, (d-g) disks with large intensity variation, (h-i) a pattern with low intensity.

The d-spacings of (1$\bar{1}$1) and (1$\bar{1}\bar{1}$) planes in real space calculated from AutoDisk are mapped in **Fig. 8a** and **8b,** respectively. The histograms of both d-spacings are plotted in **Fig. 8c**. Theoretically, the magnitude of both d-spacings should be the same in an FCC lattice. For the core-shell Pd@Pt nanoparticle, as the Pd in the core are coated with a shell of Pt atoms, the estimated d-spacings are expected to be between that of both elements, and the positions of atoms are also affected by local strain. In **Fig. 8c**, the histograms of both d-spacings follow gaussian distributions, with average magnitudes of the $d_{1\bar{1}1}$ and $d_{1\bar{1}\bar{1}}$ estimated as 2.255 Å (the green vertical line) and 2.250 Å (the blue vertical line), both in the range of the d-spacing of Pd (2.246 Å, brown dashed line) and Pt (2.2650 Å, red dashed line). Also, the estimated average lattice parameters of both d-spacings are close to $d_{111}$ of Pd, which matches with the theoretical structure.

Normal and shear strain of the Pd@Pt nanoparticle can be derived using the method in ref [14,46] and mapped in **Fig. 9**, with the average lattice of the nanoparticle as the reference. The normal strain along *X*-direction in **Fig. 9a** is almost symmetric with values between -3% and 3%. Along *Y*-axis in **Fig. 9b**, the strain is larger. Small shear strain and local lattice rotation is also identified in **Fig. 9cd**. Comparing to **Fig. 6c**, the contrast variation in the ADF image is related to the strain distribution, especially along the *X* direction. It is worth noting that on the edge of the sample **Fig. 7f-i**, AutoDisk is still able to calculate the local strain, however, with large errors due to the limited number of detectable disks in these diffraction patterns.

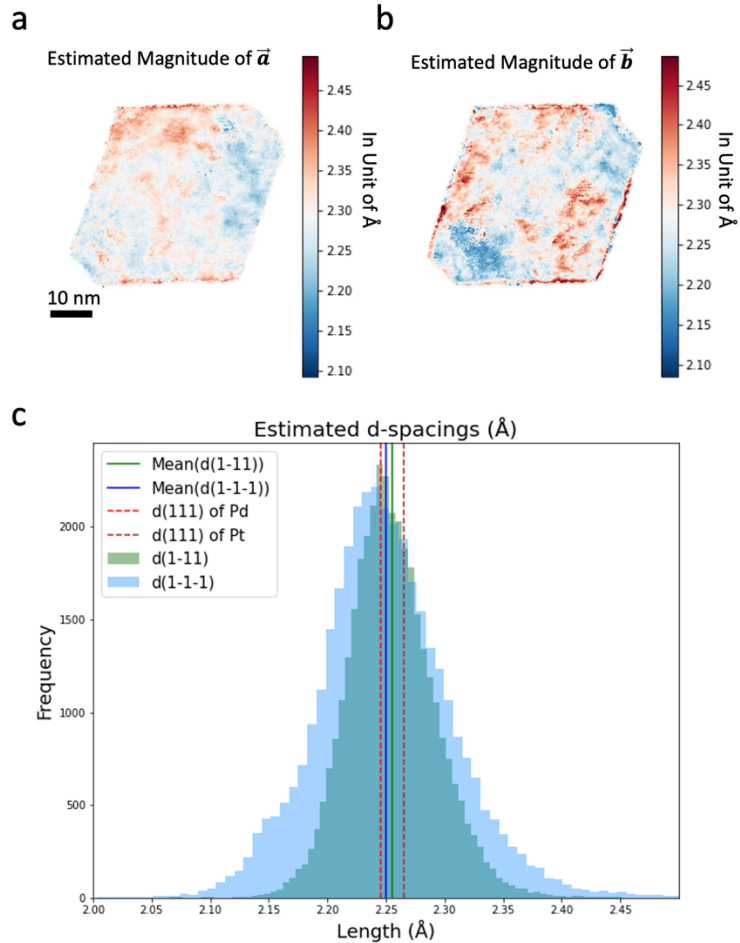

**Fig. 8.** Maps of the magnitudes of the d-spacings and their histograms. (a) The map of $d_{1\bar{1}1}$. (b) The map of d-spacings of $d_{1\bar{1}\bar{1}}$. (c) The histograms of the d-spacings.

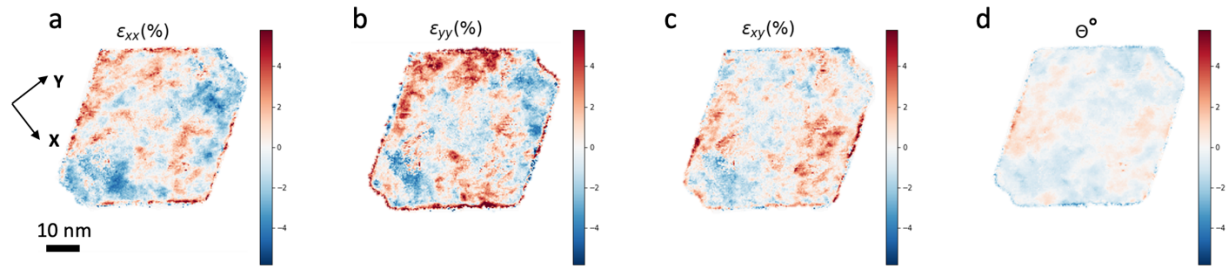

**Fig. 9.** Normal ($\varepsilon_{xx}$ and $\varepsilon_{yy}$) and shear ($\varepsilon_{xy}$) strain maps of the Pd@Pt nanoparticle in **Fig. 6**.

**Conclusions**

AutoDisk is developed here as an approach for automatic 4D-STEM data analysis. The method combines cross-correlation, blob detection, edge refinement, and lattice fitting algorithm to automate the identification of diffraction disks and measurement of local lattice parameters with a picometer accuracy. In both simulated and experimental data, AutoDisk works robustly with the presence of thickness variation and noise. Because the algorithms define and refine the reciprocal lattice without prior knowledge, it could be applied to automated processing of large data set acquired in 4D-STEM. Not only can AutoDisk be performed to lattice and strain mapping, the precise locations of diffraction disks determined by the approach can also be the basis for high-throughput characterization of phase, symmetry, orientation, and other crystallographic analysis in general.

**Acknowledgement**

This work was supported by start-up fund from the College of Engineering and Department of Materials Science and Engineering at North Carolina State University. Electron microscopy was performed at the Analytical Instrumentation Facility (AIF) at North Carolina State University, which is supported by the State of North Carolina and the National Science Foundation (award number ECCS-2025064). The AIF is a member of the North Carolina Research Triangle Nanotechnology Network (RTNN), a site in the National Nanotechnology Coordinated Infrastructure (NNCI). The authors thank Prof. Jianbo Wu from Shanghai Jiao Tong University and Prof. Hui Zhang from Zhejiang University for providing the Pd@Pt octahedral nanoparticles.

**Data Availability**

The python source code of AutoDisk and a demo are available on GitHub at https://github.com/swang59/AutoDisk_Demo. Additional raw data of the 4D-STEM experiment is available upon request to the corresponding author.

**Reference**

[1] W. Bernard, W. Rindner, H. Roth, Anisotropic Stress Effect on the Excess Current in Tunnel Diodes, Journal of Applied Physics. 35 (1964) 1860–1862. https://doi.org/10.1063/1.1713755.

[2] A.E. Romanov, T.J. Baker, S. Nakamura, J.S. Speck, ERATO/JST UCSB Group, Strain-induced polarization in wurtzite III-nitride semipolar layers, Journal of Applied Physics. 100 (2006) 023522. https://doi.org/10.1063/1.2218385.

[3] M. Chu, Y. Sun, U. Aghoram, S.E. Thompson, Strain: A Solution for Higher Carrier Mobility in Nanoscale MOSFETs, Annu. Rev. Mater. Res. 39 (2009) 203–229. https://doi.org/10.1146/annurev-matsci-082908-145312.

[4] X. Jin, A. Mano, F. Ichihashi, N. Yamamoto, Y. Takeda, High-Performance Spin-Polarized Photocathodes Using a GaAs/GaAsP Strain-Compensated Superlattice, Appl. Phys. Express. 6 (2013) 015801. https://doi.org/10.7567/APEX.6.015801.

[5] J. Li, Z. Shan, E. Ma, Elastic strain engineering for unprecedented materials properties, MRS Bull. 39 (2014) 108–114. https://doi.org/10.1557/mrs.2014.3.

[6] A.D. Krawitz, T.M. Holden, The Measurement of Residual Stresses Using Neutron Diffraction, MRS Bull. 15 (1990) 57–64. https://doi.org/10.1557/S0883769400058371.

[7] I.D. Wolf, Stress measurements in Si microelectronics devices using Raman spectroscopy, J. Raman Spectrosc. (1999) 8.

[8] E.J. Mittemeijer, U. Welzel, The "state of the art" of the diffraction analysis of crystallite size and lattice strain, Zeitschrift Für Kristallographie. 223 (2008). https://doi.org/10.1524/zkri.2008.1213.

[9] P. Strasser, S. Koh, T. Anniyev, J. Greeley, K. More, C. Yu, Z. Liu, S. Kaya, D. Nordlund, H. Ogasawara, M.F. Toney, A. Nilsson, Lattice-strain control of the activity in dealloyed core–shell fuel cell catalysts, Nature Chem. 2 (2010) 454–460. https://doi.org/10.1038/nchem.623.

[10] T. Nilsson Pingel, M. Jørgensen, A.B. Yankovich, H. Grönbeck, E. Olsson, Influence of atomic site-specific strain on catalytic activity of supported nanoparticles, Nat Commun. 9 (2018) 2722. https://doi.org/10.1038/s41467-018-05055-1.

[11] R. Yuan, J. Zhang, J.-M. Zuo, Lattice strain mapping using circular Hough transform for electron diffraction disk detection, Ultramicroscopy. 207 (2019) 112837. https://doi.org/10.1016/j.ultramic.2019.112837.

[12] R. Wittmann, C. Parzinger, D. Gerthsen, Quantitative determination of lattice parameters from CBED patterns: accuracy and performance, Ultramicroscopy. 70 (1998) 145–159. https://doi.org/10.1016/S0304-3991(97)00107-1.

[13] M.J. Hÿtch, E. Snoeck, R. Kilaas, Quantitative measurement of displacement and strain fields from HREM micrographs, Ultramicroscopy. 74 (1998) 131–146. https://doi.org/10.1016/S0304-3991(98)00035-7.

[14] A. Béché, J.L. Rouvière, L. Clément, J.M. Hartmann, Improved precision in strain measurement using nanobeam electron diffraction, Appl. Phys. Lett. 95 (2009) 123114. https://doi.org/10.1063/1.3224886.

[15] C. Ophus, P. Ercius, M. Sarahan, C. Czarnik, J. Ciston, Recording and Using 4D-STEM Datasets in Materials Science, Microsc Microanal. 20 (2014) 62–63. https://doi.org/10.1017/S1431927614002037.

[16] C. Ophus, Four-Dimensional Scanning Transmission Electron Microscopy (4D-STEM): From Scanning Nanodiffraction to Ptychography and Beyond, Microsc Microanal. 25 (2019) 563–582. https://doi.org/10.1017/S1431927619000497.

[17] H. Yang, L. Jones, H. Ryll, M. Simson, H. Soltau, Y. Kondo, R. Sagawa, H. Banba, I. MacLaren, P.D. Nellist, 4D STEM: High efficiency phase contrast imaging using a fast

pixelated detector, J. Phys.: Conf. Ser. 644 (2015) 012032. https://doi.org/10.1088/1742-6596/644/1/012032.

[18] D.B. Williams, C.B. Carter, Transmission electron microscopy: a textbook for materials science, 2nd ed, Springer, New York, 2008.

[19] A. Béché, J.L. Rouvière, J.P. Barnes, D. Cooper, Strain measurement at the nanoscale: Comparison between convergent beam electron diffraction, nano-beam electron diffraction, high resolution imaging and dark field electron holography, Ultramicroscopy. 131 (2013) 10–23. https://doi.org/10.1016/j.ultramic.2013.03.014.

[20] J.-M. Zuo, A.B. Shah, H. Kim, Y. Meng, W. Gao, J.-L. Rouviére, Lattice and strain analysis of atomic resolution Z-contrast images based on template matching, Ultramicroscopy. 136 (2014) 50–60. https://doi.org/10.1016/j.ultramic.2013.07.018.

[21] P.L. Galindo, S. Kret, A.M. Sanchez, J.-Y. Laval, A. Yáñez, J. Pizarro, E. Guerrero, T. Ben, S.I. Molina, The Peak Pairs algorithm for strain mapping from HRTEM images, Ultramicroscopy. 107 (2007) 1186–1193. https://doi.org/10.1016/j.ultramic.2007.01.019.

[22] K. Usuda, T. Mizuno, T. Tezuka, N. Sugiyama, Y. Moriyama, S. Nakaharai, S. Takagi, Strain relaxation of strained-Si layers on SiGe-on-insulator (SGOI) structures after mesa isolation, Applied Surface Science. 224 (2004) 113–116. https://doi.org/10.1016/j.apsusc.2003.11.058.

[23] T.C. Pekin, C. Gammer, J. Ciston, A.M. Minor, C. Ophus, Optimizing disk registration algorithms for nanobeam electron diffraction strain mapping, Ultramicroscopy. 176 (2017) 170–176. https://doi.org/10.1016/j.ultramic.2016.12.021.

[24] C. Gammer, C. Ophus, T.C. Pekin, J. Eckert, A.M. Minor, Local nanoscale strain mapping of a metallic glass during *in situ* testing, Appl. Phys. Lett. 112 (2018) 171905. https://doi.org/10.1063/1.5025686.

[25] T.C. Pekin, C. Gammer, J. Ciston, C. Ophus, A.M. Minor, In situ nanobeam electron diffraction strain mapping of planar slip in stainless steel, Scripta Materialia. 146 (2018) 87–90. https://doi.org/10.1016/j.scriptamat.2017.11.005.

[26] Y. Han, K. Nguyen, M. Cao, P. Cueva, S. Xie, M.W. Tate, P. Purohit, S.M. Gruner, J. Park, D.A. Muller, Strain Mapping of Two-Dimensional Heterostructures with Subpicometer Precision, Nano Lett. 18 (2018) 3746–3751. https://doi.org/10.1021/acs.nanolett.8b00952.

[27] J.M. Cowley, A.F. Moodie, The scattering of electrons by atoms and crystals. I. A new theoretical approach, Acta Cryst. 10 (1957) 609–619. https://doi.org/10.1107/S0365110X57002194.

[28] K. Müller, A. Rosenauer, M. Schowalter, J. Zweck, R. Fritz, K. Volz, Strain Measurement in Semiconductor Heterostructures by Scanning Transmission Electron Microscopy, Microsc Microanal. 18 (2012) 995–1009. https://doi.org/10.1017/S1431927612001274.

[29] V.B. Ozdol, C. Gammer, X.G. Jin, P. Ercius, C. Ophus, J. Ciston, A.M. Minor, Strain mapping at nanometer resolution using advanced nano-beam electron diffraction, Appl. Phys. Lett. 106 (2015) 253107. https://doi.org/10.1063/1.4922994.

[30] C. Mahr, K. Müller-Caspary, T. Grieb, M. Schowalter, T. Mehrtens, F.F. Krause, D. Zillmann, A. Rosenauer, Theoretical study of precision and accuracy of strain analysis by nano-beam electron diffraction, Ultramicroscopy. 158 (2015) 38–48. https://doi.org/10.1016/j.ultramic.2015.06.011.

[31] B.H. Savitzky, L. Hughes, K.C. Bustillo, H.D. Deng, N.L. Jin, E.G. Lomeli, W.C. Chueh, P. Herring, A. Minor, C. Ophus, py4DSTEM: Open Source Software for 4D-STEM Data

Analysis, Microsc Microanal. 25 (2019) 124–125. https://doi.org/10.1017/S1431927619001351.

[32] E. Padgett, M.E. Holtz, P. Cueva, Y.-T. Shao, E. Langenberg, D.G. Schlom, D.A. Muller, The exit-wave power-cepstrum transform for scanning nanobeam electron diffraction: robust strain mapping at subnanometer resolution and subpicometer precision, Ultramicroscopy. 214 (2020) 112994. https://doi.org/10.1016/j.ultramic.2020.112994.

[33] S.E. Zeltmann, A. Müller, K.C. Bustillo, B. Savitzky, L. Hughes, A.M. Minor, C. Ophus, Patterned probes for high precision 4D-STEM bragg measurements, Ultramicroscopy. 209 (2020) 112890. https://doi.org/10.1016/j.ultramic.2019.112890.

[34] M. Klinger, M. Němec, L. Polívka, V. Gärtnerová, A. Jäger, Automated CBED processing: Sample thickness estimation based on analysis of zone-axis CBED pattern, Ultramicroscopy. 150 (2015) 88–95. https://doi.org/10.1016/j.ultramic.2014.12.006.

[35] J.M. Zuo, J.C. Mabon, Web-Based Electron Microscopy Application Software: Web-EMAPS, Microsc Microanal. 10 (2004) 1000–1001. https://doi.org/10.1017/S1431927604884319.

[36] M.W. Tate, P. Purohit, D. Chamberlain, K.X. Nguyen, R. Hovden, C.S. Chang, P. Deb, E. Turgut, J.T. Heron, D.G. Schlom, D.C. Ralph, G.D. Fuchs, K.S. Shanks, H.T. Philipp, D.A. Muller, S.M. Gruner, High Dynamic Range Pixel Array Detector for Scanning Transmission Electron Microscopy, Microsc Microanal. 22 (2016) 237–249. https://doi.org/10.1017/S1431927615015664.

[37] J.C.H. Spence, J.M. Zuo, A Brief History of Electron Microdiffraction, in: Electron Microdiffraction, Springer US, Boston, MA, 1992: pp. 1–5. https://doi.org/10.1007/978-1-4899-2353-0_1.

[38] C.R. Harris, K.J. Millman, S.J. van der Walt, R. Gommers, P. Virtanen, D. Cournapeau, E. Wieser, J. Taylor, S. Berg, N.J. Smith, R. Kern, M. Picus, S. Hoyer, M.H. van Kerkwijk, M. Brett, A. Haldane, J.F. del Río, M. Wiebe, P. Peterson, P. Gérard-Marchant, K. Sheppard, T. Reddy, W. Weckesser, H. Abbasi, C. Gohlke, T.E. Oliphant, Array Programming with NumPy, Nature. 585 (2020) 357–362. https://doi.org/10.1038/s41586-020-2649-2.

[39] G.R. Bradski, A. Kaehler, Learning OpenCV: computer vision with the OpenCV library, 1. ed., [Nachdr.], O'Reilly, Beijing, 2011.

[40] J.D. Hunter, Matplotlib: A 2D Graphics Environment, Comput. Sci. Eng. 9 (2007) 90–95. https://doi.org/10.1109/MCSE.2007.55.

[41] S. van der Walt, J.L. Schönberger, J. Nunez-Iglesias, F. Boulogne, J.D. Warner, N. Yager, E. Gouillart, T. Yu, scikit-image: image processing in Python, PeerJ. 2 (2014) e453. https://doi.org/10.7717/peerj.453.

[42] T. Lindeberg, Feature Detection with Automatic Scale Selection, (1998) 38.

[43] H. Kong, H.C. Akakin, S.E. Sarma, A Generalized Laplacian of Gaussian Filter for Blob Detection and Its Applications, IEEE Trans. Cybern. 43 (2013) 1719–1733. https://doi.org/10.1109/TSMCB.2012.2228639.

[44] T. Grieb, F.F. Krause, C. Mahr, D. Zillmann, K. Müller-Caspary, M. Schowalter, A. Rosenauer, Optimization of NBED simulations for disc-detection measurements, Ultramicroscopy. 181 (2017) 50–60. https://doi.org/10.1016/j.ultramic.2017.04.015.

[45] J.W. Tukey, Exploratory data analysis, Reading, MA, 1977.

[46] J.L. Rouvière, E. Sarigiannidou, Theoretical discussions on the geometrical phase analysis, Ultramicroscopy. 106 (2005) 1–17. https://doi.org/10.1016/j.ultramic.2005.06.001.